\documentstyle[twoside,fleqn,espcrc2,epsfig]{article}

\newcommand{\AmS}{{\protect\the\textfont2
  A\kern-.1667em\lower.5ex\hbox{M}\kern-.125emS}}
\newcommand{\be}{\begin{equation}}
\newcommand{\ee}{\end{equation}}
\newcommand{\gtsima}{$\; \buildrel > \over \sim \;$}
\newcommand{\ltsima}{$\; \buildrel < \over \sim \;$}
\newcommand{\simlt}{\lower.5ex\hbox{\ltsima}}
\newcommand{\simgt}{\lower.5ex\hbox{\gtsima}}
\title{High Energy Phenomena in Clusters of Galaxies}
\author{P. Blasi \address{Department of Astronomy \& Astrophysics, 
and\\
E. Fermi Institute, The University of Chicago,\\
5640 S. Ellis Av., Chicago, IL 60637}
and
S. Colafrancesco\address{Osservatorio Astronomico di Roma,\\
Via dell'Osservatorio 2, I-00040, Monteporzio (ITALY)}}

\begin{document}

\begin{abstract}
Several phenomena in high energy astrophysics have been recently 
related
to clusters of galaxies and to cosmic ray interactions occurring 
inside these structures. 
In many of these phenomena the observable effects 
depend on the energy density of 
cosmic rays confined in the Intra Cluster (IC) 
medium, which is a poorly known quantity.
We propose here that useful indications about this quantity can be obtained 
from present and future observations of galaxy clusters 
in the radio and hard X-ray frequency ranges.

\end{abstract}

\maketitle

\section{Introduction}

Clusters of galaxies are the largest gravitationally bound structures
in the universe.
They are exceptionally useful laboratories for {\it cosmology} and 
{\it high energy astrophysics}.

From the cosmological side these structures probe:

\noindent
{\it i)}  the amplitude and shape of the primordial fluctuation 
spectrum, because the mass (or temperature) function
of rich clusters is strongly
sensitive to the value of the fluctuation power-spectrum, 
$\sigma(M,z)$. Thus it is possible to measure $\sigma(M,z)$ on the
cluster scales fitting the mass or temperature function to the local
data (see, e.g., \cite{cmv97} \cite{neta} and references therein);

\noindent
{\it ii)}  the evolution of baryons, both condensed in the form of galaxies
and diffuse in the form of IC gas, which is abundantly
present within the cluster potential wells.
In fact, galaxies are responsible for the chemical
enrichment of the IC gas at a level $Fe/H \simgt 0.3$ of the solar
value, during their life-cycles.
Thus,  high-quality X-ray
spectral observations of galaxy clusters allow to study in
details the physical state of the IC gas and the feedbacks
from galaxy evolution \cite{c97};

\noindent 
{\it iii)} the overall structure of the universe, using the
possibility to measure directly $H_0$ and $\Omega_0$ through X-ray and
Sunyaev-Zel'dovich effect (in the radio and sub-mm bands) 
cluster observations \cite{yr}.

Clusters of galaxies are also relevant for high energy 
phenomena in large scale structures because they can be regarded as 
the {\it largest particle accelerators in the sky}.
In the following, we will discuss some aspects in which
galaxy clusters can yield important insights for high energy
astrophysical phenomena.
 
\section{Clusters of galaxies as $\gamma$-ray sources}
There are different types and sites of particle acceleration in
galaxy clusters from which $\gamma$-rays and neutrinos can be produced
through the following channels:
$$
p+p\to\pi^0+X, ~~~~~~\pi^0 \to \gamma+\gamma
$$
$$
p+p\to\pi^{\pm}+X, ~~~~~\pi^{\pm}\to \mu^{\pm} 
\nu_{\mu}(\bar{\nu}_{\mu}), 
$$
\be
~~~~~~~~~~~~~~~~~\mu^{\pm}\to e^{\pm} + 
\bar{\nu}_{\mu}(\nu_{\mu})
+ \nu_e (\bar{\nu}_e) ~.
\ee
The possible sites are \cite{bbp} \cite{cb97}: 
{\it i)} normal galaxies with low cosmic ray 
(CR) emitted powers, $L_{CR} \simlt
10^{43}$ erg/s; 
{\it ii)} active galaxies (like radio-galaxies living in the cluster
cores or AGNs) with moderate powers, $L_{CR} \sim
10^{44}$ erg/s; 
{\it iii)} accretion and/or merging shocks which are produced during
the cluster collapse and virialization (see Fig.1) 
and can produce
large powers, $L_{CR} \simgt 2 \cdot 10^{44}$ erg/s, through first
order Fermi acceleration mechanism.
\begin{figure}[thb]
 \begin{center}
  \mbox{\epsfig{file=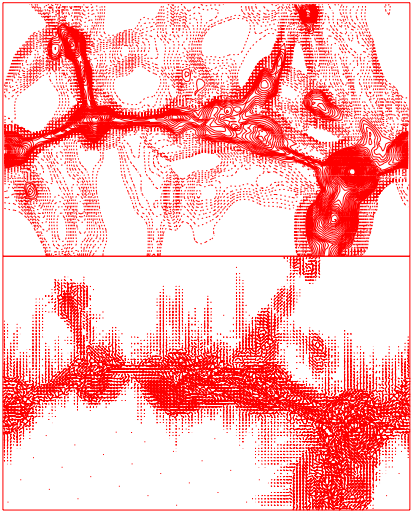,width=4.cm}}
  \caption{\em {Shocks in large scale structures from
N-body simulations {\it [7]} (courtesy of D.Ryu) 
in which are shown the
distribution of the baryonic matter (upper panel) and of the magnetic
field (lower panel).
Clusters of galaxies are found as the
knots of the large scale matter condensations.
}}
 \end{center}
\end{figure}
In the $pp$ collisions giving rise to $\gamma$-ray production, bullet
protons are supplied by the previous particle accelerators found
within galaxy clusters, while the target protons are those of the
hot ($T \sim 10^7\div 10^8$ K), diffuse ($n_e \sim 10^{-3}~cm^{-3}$) and
metal enriched IC gas (see \cite{sar88} for a review) which is
responsible for the cluster X-ray thermal emission.
Thus, a correlation is expected between the X-ray (thermal)
power: 
\be
L_X \propto n_0^2 T^{1/2} r_c^3 ~,
\ee
and the $\gamma$-ray (non-thermal) power:
\be
Q_i(E) \propto Y_i \sigma_{pp} c n_0 Q_p(E) r_c^2 ~,
\ee
where $n_0$ is the central density of the IC gas, $T$ its temperature
and $r_c$ the core radius of a cluster with density profile $n(r) = n_0
(1+x^2)^{-3\beta/2}$, with $x\equiv r/r_c$ and $\beta \sim 0.6\div 
0.8$ (see \cite{cb97} for details).
The correlation among the X-ray and $\gamma$-ray fluxes 
(see Fig.2) reads:
\be
F_{\gamma} \propto F_X \bigg( n_0 r_c T^{1/2} \bigg)^{-1}~.
\ee
\begin{figure}[thb]
 \begin{center}
  \mbox{\epsfig{file=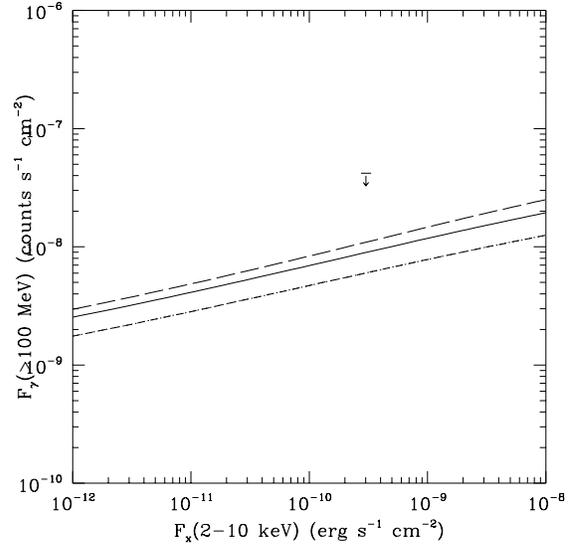,width=7.5cm}}
  \caption{\em {Gamma ray fluxes for clusters at  $z =0.023$ in
different cosmological scenarios:
CDM ($\Omega_0=1$; dashed curve), CDM$+\Lambda$ 
($\Omega_0=0.4$; continuous curve), open CDM ($\Omega_0=0.3$; 
long-dashes curve)
and mixed DM ($\Omega_0=1$, $\Omega_{\nu}=0.3$; dotted curve).
The arrow indicates the EGRET upper
limit for A1656 (Coma).}}
 \end{center}
\end{figure}

In the previous formulae, $Q_p(E)$ is the CR spectrum which 
determines $L_{CR}$ and consequently the cluster $\gamma$-ray
emissivity, but it is a
poorly known quantity in clusters of galaxies and in large scale
structures in general.
Assuming $L_{CR} \approx 10^{44}$ erg/s (as produced by accretion
shocks and/or active galaxies) we predicted that the contribution of
galaxy clusters to the diffuse gamma ray background is $\simlt 3 \%$,
quite independently of the assumed cosmological scenario and on
cluster evolution (see Fig.3).
\begin{figure}[thb]
 \begin{center}
  \mbox{\epsfig{file=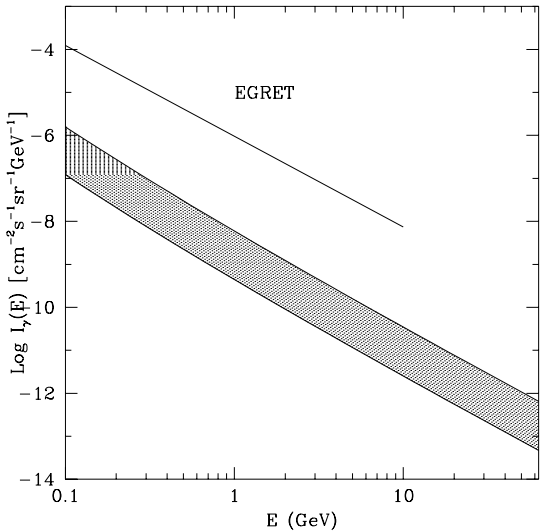,width=7.5cm}}
  \caption{\em {The expected DGRB from clusters,
considering various sources of theoretical uncertainties in the cluster
modelling. A flat CDM ($\Omega_0=1; h=0.5; n=1$) model 
is considered here.
The thick solid line show the best fit to the EGRET data
{\it [9]}.}}
 \end{center}
\end{figure}
This result (apparently not relevant for the discussion on the origin
of the DGRB) is indeed relevant because it shows that galaxy clusters 
cannot overcome
the fraction of the DGRB left available from other 
(resolved or truly diffuse) sources summing up to $\sim 80
\div 95 \%$ (see discussion in \cite{cb97}).
Larger fractions of the DGRB contributed by galaxy clusters would
give sensitive constraints to the current scenarios for structure formation.

Note, however, that these results rely on the 
(reasonable and motivated) value assumed for 
the CR power in clusters,  $L_{CR} \approx 10^{44}$ erg/s, which 
depends on the CR spectrum $Q_p(E)$.
So far, there are only heuristic derivations of such a crucial
quantity.
Nonetheless, in the following we will propose an operative procedure
to obtain quantitative indications on the CR spectrum in galaxy
clusters.

\section{Non-thermal emission from secondary electrons in
galaxy clusters}

Most of the cosmic rays produced in galaxy
clusters are confined in the IC medium for times
larger than the age of the universe \cite{bbp} \cite{cb97}.
However, the maximum energy for which this 
confinement is still efficient depends strongly on the 
adopted diffusion model,
so that for very extreme choices of the 
parameters, confinement can be
limited to low energies. 
Here we shall consider the two limiting cases of complete 
confinement and no confinement in order to outline
the general picture of the CR diffusion in clusters. 
During the lifetime of the CR in the
IC medium, they interact with the hot IC
protons and produce secondary particles through $pp$ collisions (see
eq.1).
In such collisions charged pions
are also produced and they produce a flux of neutrinos (see 
\cite{bbp} \cite{cb97})
and electrons (here we will refer to both electrons and positrons
as 'electrons', being this distinction irrelevant for our purposes).

The global normalization of the gamma or neutrino spectra 
obtained in
previous works depends on a poorly known quantity, namely the 
energy density of CRs in clusters. 
Here we  propose that useful indications
on the CR energy density in clusters can be obtained by the radio
measurements already existing and by  X-ray measurements 
in the hard part of the spectrum, 
that will be available in the next future.
In our model, radio and non-thermal X-ray 
emissions are produced
by relativistic electrons through synchrotron and inverse compton~
scattering (ICS) off the microwave background photons, respectively.

We assume here a CR spectrum of the form:
\begin{equation}
Q_p(E)=\frac{L_p}{E_0^2} (\gamma_g-1)(\gamma_g-2) \left(~
\frac{E}{E_0}\right)^{-\gamma_g},
\label{eq:prod}
\end{equation}
\noindent
where $E_0\sim 1~GeV$ and $L_p$ is the cluster total CR luminosity.
The rate of production of secondary
electrons from $pp$ collisions is \cite{bc97}:
$$
Q_e(E_e)= 36 \frac{I(\gamma_g) (\gamma_g-1)(\gamma_g-2)}
{\gamma_g (\gamma_g+2) (\gamma_g+3)} L_p n_H c \sigma_0 t_0
$$
\begin{equation}
\frac{m_{\pi}^2}{m_{\pi}^2-m_{\mu}^2} E_e^{-\gamma_g},
\label{eq:qe_conf}
\end{equation}
\noindent
in the case of complete CR confinement.
For the case of no confinement we obtain:
$$
Q_e(E_e)= 36 \frac{I(\gamma_g,\eta) (\gamma_g-1)(\gamma_g-2)}
{(\gamma_g+\eta) (\gamma_g+\eta+2) (\gamma_g+\eta+3)}~
$$
\begin{equation}
L_p n_H c \sigma_0
\frac{R_{cl}^2}{6 D_0}
\frac{m_{\pi}^2}{m_{\pi}^2-m_{\mu}^2} E_e^{-\gamma_g-\eta}
\label{eq:qe_noconf}
\end{equation}
\noindent
Here $n_H$ is the average gas density in~
clusters, $\sigma_0=3.2\times 10^{-26}cm^2$, $t_0=2.06\times 10^{17}
h^{-1}~s$ is the age of universe and $m_{\pi}$ and $m_{\mu}$ are the pion
and muon masses. 
The diffusion coefficient can been written 
as 
$D(E)=D_0 E^{\eta}$. The functions $I(\gamma_g)$ and 
$I(\gamma_g,\eta)$ are
defined as $I(\gamma_g)=\int_0^1 dx f_{\pi}(x) x^{\gamma_g-2}$ and
$I(\gamma_g,\eta)=\int_0^1 dx f_{\pi}(x) x^{\gamma_g+\eta-2}$, where~
$f_{\pi}(x)=1.34 (1-x)^{3.5}+e^{-18x}$, is a scaling function entering~
the calculation of the secondary electrons from the decay~
of charged pions (see \cite{bc97}). 
In the following we use
$\gamma_g=2.4$, checking at the end how the results change for
$\gamma_g=2.1$. 

The diffusion coefficient enters effectively
in the calculation of the case in which no confinement is realized. 
For the case of a Kolmogorov spectrum of fluctuations, it
can be easily written in terms of the average magnetic field $B$, the~
coherence scale $l_c$ and the CR energy $E$ (see \cite{bc97}): 
$$D(E)\simeq$$
\begin{equation}
3\times 10^{31} E^{1/3} B_{\mu}^{-1/3}~
\left( \frac{l_c}{430~kpc}\right)^{2/3} ~ cm^2 ~ s^{-1},
\label{eq:coef_dif}
\end{equation}
\noindent
where $E$ is the CR energy in GeV and $B_{\mu}$ is the magnetic field
in $\mu G$. 
The coherence scale, $l_c$, of the magnetic field is a quite
uncertain parameter: we assumed for it an upper value, $430$ kpc,
corresponding
to the case in which the magnetic field lines are stretched by the bulk
motion of the galaxies in the clusters on a typical scale equal to the
average galaxy separation.

At the electron energies we are interested in, the time for energy losses
is smaller than the typical diffusion time, so that the equilibrium
electron spectrum is $n_e(E_e)\propto Q_e(E_e) \tau_e(E_e)$, where
$\tau_e(E_e)=E_e/(dE_e/dt)$ is the time scale for losses.
The details of the calculation of the radio and X-ray emission from
synchrotron and ICS are given in \cite{bc97}; here we shall limit 
ourselves to outline the results obtained in different cases.

In the case of complete CR confinement, the radio power spectrum reads:
$$P_r(\nu) =$$
\begin{equation}
1.7\times 10^{42} L_{44} n_3 h_{75}^{-1} \nu(Hz)^{-1.2}
~erg~s^{-1}~Hz^{-1},
\label{eq:prcon1}
\end{equation}
\noindent
with $B=1~\mu G$ and $\gamma_g=2.4$. 
Here $L_{44}$ is the cluster CR luminosity in
units of $10^{44}~erg/s$, $n_3$ is the gas density in units of~
$10^{-3}cm^{-3}$ and $h_{75}$ is the Hubble constant in units of
$75~km/(s~Mpc)$.
With the same parameter choice, the X-ray spectrum reads:
$$I_X(E_X) =$$
\begin{equation}
5.3\times 10^{50} L_{44} n_3 h_{75}^{-1} E_X(keV)^{-2.2}
s^{-1}~keV^{-1}.
\label{eq:pxcon1}
\end{equation}
While the X-ray flux is very weakly dependent on the magnetic field,
the radio flux changes strongly with $B$. If we take
$B\approx 0.1~\mu G$ eq.(\ref{eq:prcon1}) yields;
$$P_r(\nu) =$$
\begin{equation}
1.2\times 10^{40} L_{44} n_3 h_{75}^{-1} \nu(Hz)^{-1.2}
~erg~s^{-1}~Hz^{-1}.
\label{eq:prcon2}
\end{equation}
This fact has a relevant implication: 
an accurate measurement of the flux in the hard X-ray region 
of the cluster spectra
can fix the product $L_{44} n_3$ so that, in
turn,  a measure of $P_r(\nu)$ can give informations 
on the value of $B$.

In the case of {\it no confinement} we obtain for the radio
power spectrum:
$$P_r(\nu) =$$
\begin{equation}
5.6\times 10^{42} L_{44} n_3 h_{75}^{-1} \nu(Hz)^{-1.365}
~erg~s^{-1}~Hz^{-1},
\label{eq:prcon3}
\end{equation}
\noindent
and for the X-ray spectrum:
$$I_X(E_X) =$$
\begin{equation}
1.4\times 10^{50} L_{44} n_3 h_{75}^{-1} E_X(keV)^{-2.365}
s^{-1}~keV^{-1}.
\label{eq:pxcon2}
\end{equation}
Again $I_X(E_X)$ does not change appreciably with the magnetic field,
while $P_r(\nu)$ does. For $B=0.1~\mu G$ we obtain:
$$P_r(\nu)=$$\par
\begin{equation}
1.3\times 10^{40} L_{44} n_3 h_{75}^{-1} \nu(Hz)^{-1.365}
~erg~s^{-1}~Hz^{-1}.
\label{eq:prcon4}
\end{equation}

\section{A future outlook}
At present, there are only a few observations of cluster radio halo
spectra, the most detailed one referring to the Coma cluster 
\cite{fg97}.
Based on the available observations, we can only put 
weak constraints to the cluster CR
energy density,
$\omega_{CR}^{cl}\simlt 10^{-2}~eV/cm^3$ 
(confinement with $B=1\mu G$) or
$\omega_{CR}^{cl}\simlt 1~eV/cm^3$ (confinement with $B=0.1\mu G$).
Moreover, note that the EGRET upper limit on the $\gamma$-ray emission
from Coma (see Fig.2) yields
$\omega^{cl}_{CR}\simlt 0.1~eV/cm^3$. 

From this first comparison of our results with the
available data, it is
evident the extreme importance of a definite detection of non thermal 
X-ray tail in the spectra of galaxy clusters.
In fact,
such X-ray spectra are very weakly dependent on the 
value of the magnetic field, so that X-rays can in principle fix 
the value of 
$L_{44}n_3$, provided that the radio halos observed in the same 
clusters are due to the same secondary electrons. 
Then, in turn, radio observations can be used to give 
reliable indications concerning both 
the amplitude and the power spectrum of the magnetic field in clusters.

The available
observations at $E \geq 20$ keV from Coma can only provide 
upper limits to the hard, non-thermal X-ray  tails \cite{rug94}.
%(Rephaeli, Ulmer and Gruber, 1994). 
In our model, these  limits yield
$L_{44} n_3 \simlt  60\div 600$ where the two values 
refer to the extreme
cases of confinement and no confinement, respectively.
These give $L_{44} \simlt 1.4 \cdot 10^2 \div 1.4 \cdot 10^3$  
for an average IC gas density in Coma of $n_3 \approx 0.43$.
The previous limits  are very weak and do not allow,
at the moment,
neither to distinguish between the two regimes nor to strongly
constrain the CR energy density in Coma.
The results obtained for the case of an electron 
production spectrum with index $\gamma_g=2.4$, do not change
appreciably if we use $\gamma_g=2.1$, which is somewhat close to a
lower limit for this slope if the parent CRs are assumed to be produced
by acceleration at supernovae shocks.

Beyond the details of the calculations we presented here,
the possible detection of hard X-ray tails in the spectra 
of galaxy clusters will provide new insights for the role of
high-E phenomena in large scale structures.
This is at hand in the hard X-ray region
with the data on galaxy clusters achievable with the 
SAX satellite, which is now operating. 
In fact, the PDS instrument on board SAX 
has a limiting flux $\simgt 10^{-11} 
erg/cm^2 s$, with $\Delta E \sim 15 \% $ (at $60$ keV) 
and a broad band, $\Delta E 
\sim 15 \div 300$ keV, response which makes it possible to detect 
hard X-ray 
tails in the spectra of several nearby clusters like Coma, A2163 and
A2199.

Besides the analysis of the two extreme cases of CR confinement we
presented here,
the detection of a possible steepening in the X-ray tail spectra can
suggest (see sect.3)
that CR begin to be unconfined above some threshold energy, while 
no effect should be present in the radio spectrum. A more detailed
discussion of this effect will be reported in a forthcoming paper
\cite{bc98}.
%(Blasi and Colafrancesco, 1997).

Thus, in the light of the present results, 
better constraints on the CR energy 
density in the IC medium are expected to become available 
in the next future, 
together with a better insight on the structure of the 
IC magnetic fields.


\begin{thebibliography}{9}

\bibitem{cmv97} S. Colafrancesco, P. Mazzotta \& N. Vittorio 1997, 
ApJ, 488, 566

\bibitem{neta} N. Bahcall 1997, preprint astro-ph/9711062

\bibitem{c97} S. Colafrancesco 1998, preprint 

\bibitem{yr} Y. Rephaeli 1995, ARA\&A, 33, 541

\bibitem{bbp}
V.S. Berezinsky, P. Blasi \& V.S. Ptuskin 1997, ApJ, 487,  529

\bibitem{cb97}
S. Colafrancesco \& P. Blasi, Astroparticle Physics in press,
preprint astro-ph/9804262

\bibitem{rk97} D.Ryu \& H. Kang 1997, preprint astro/ph-9702055 

\bibitem{sar88} C. Sarazin 1988, {\it X-Ray Emission from Clusters of
Galaxies}, Cambridge Univ. Press

%\bibitem{dar}
%A. Dar and N.J. Shaviv, Phys. Rev. Lett., 75 (1995) 3052.

\bibitem{owz} J.L. Osborne et al. 1994, J. Phys. G., 20, 1089

\bibitem{bc97}
P. Blasi \& S. Colafrancesco 1998, submitted to Astroparticle Physics

\bibitem{bc98}
P. Blasi \& S. Colafrancesco 1998, in preparation

\bibitem{fg97} 
%L. Feretti \& G. Giovannini 1997, in {\it Extragalactic
%Radio Sources}, Eds. R. Ekers et al.
K.T. Kim et al. 1990, ApJ, 355, 29

\bibitem{rug94} Y. Rephaeli, M. Ulmer \& D. Gruber 1994, ApJ, 429, 554

\end{thebibliography}
\end{document}